\documentclass[a4paper]{article}

\usepackage{INTERSPEECH2022}

\usepackage{subfigure}
\usepackage{cite}
\usepackage{multirow}
\usepackage{booktabs}
\usepackage[urlcolor=blue]{hyperref}
\title{Towards Multi-Scale Speaking Style Modelling with Hierarchical Context Information for Mandarin Speech Synthesis}
\name{Shun Lei$^{1\ddagger}$\thanks{$^{\ddagger}$Work conducted when the first author was intern at Huya Inc.}, Yixuan Zhou$^{1\dagger}$, Liyang Chen$^1$, Jiankun Hu$^2$, Zhiyong Wu$^{1,4*}$\thanks{$^{\dagger}$ Equal contribution. $^{*}$ Corresponding author.},  Shiyin Kang$^3$, Helen Meng$^4$}
\address{
    $^1$ Shenzhen International Graduate School, Tsinghua University, Shenzhen, China\\
    $^2$ Huya Inc., Guangzhou, China\quad$^3$ XVerse Inc., Shenzhen, China\\
    $^4$ The Chinese University of Hong Kong, Hong Kong SAR, China
}
\email{\{leis21, zhouyx20, cly21\}$@$mails.tsinghua.edu.cn, zywu$@$sz.tsinghua.edu.cn
}

\begin{document}

\maketitle
\begin{abstract}
Previous works on expressive speech synthesis focus on modelling the mono-scale style embedding from the current sentence or context, but the multi-scale nature of speaking style in human speech is neglected.
In this paper, we propose a multi-scale speaking style modelling method to 
capture and predict multi-scale speaking style for improving the naturalness and expressiveness of synthetic speech.
A multi-scale extractor is proposed to extract speaking style embeddings at three different levels from the ground-truth speech, and explicitly guide the training of a multi-scale style predictor based on hierarchical context information.
Both objective and subjective evaluations on a Mandarin audiobooks dataset demonstrate that our proposed method can significantly improve the naturalness and expressiveness of the synthesized speech\footnote{Speech samples: \href{https://thuhcsi.github.io/interspeech2022-msc-tts}{https://thuhcsi.github.io/interspeech2022-msc-tts}}.
\end{abstract}
\section{Introduction}
Text-to-speech (TTS) aims to generate 
intelligible and natural speech from text. 
With the development of deep learning, TTS models are now embodied with the ability to synthesize high-quality speech with a neutral speaking style \cite{tacotron2, deepvoice3, fastspeech2}.
However, the speaking style with limited expressiveness remains a clear gap between synthesized speeches and human recordings, which blocks the development of speech synthesis technology in many application scenarios such as audiobooks, podcasts, and voice assistants.
Therefore, how to model expressive speaking style is a hot research topic in academia and industry recently.

One of the general approaches is
to extract the speaking style representation from given reference audio \cite{reference, gst, VAEGST, wu2019end},
which synthesizes speech conditioned on the extracted representation.
Compared with the reference audio-based methods, another line that directly predicting the speaking style from text without auxiliary inputs is more practical and flexible.
The text-predicted global style token (TP-GST) model \cite{TPGST} is proposed to predict the global-level style representation from text alone.
Benefiting from the great semantic representation ability of the pre-trained language models, such as bidirectional encoder representations from Transformer (BERT) \cite{bert}, 
text representations derived from the pre-trained language model have been used to predict speaking style and shown gains in performance \cite{berttacotron, bertemb}.

Some early text-predicted methods concentrate only on the current sentence, which fails to capture the style information 
influenced by the different context of neighbor sentences \cite{survey, longformevaluationg}.
To avoid this problem, \cite{crossutterance} proposes to use the neighbor sentences to improve the prosody generation.
Our preliminary work \cite{proposed} utilizes the hierarchical context encoder (HCE) to further consider the hierarchical structure of context and predicts the global-scale speaking style
in an explicit way.
These studies demonstrate that taking a wider range of contextual information into account is helpful for expressive speech synthesis.

However, HCE still suffers from the absence of local-scale style modeling (e.g., intonation, rhythm, stress).
To model and control local prosodic variations in speech, some previous works attempt to predict finer-grained speaking styles from text, such as word level \cite{wsv, ren2022prosospeech} and phoneme level \cite{finegrained, du2021mixture}.
It is more widely accepted that the style expressions of human speech are multi-scale in nature \cite{selkirk1986derived, liberman1977stress}, where the global-scale style is usually observed as emotion and the local-scale is more close to the prosody variation \cite{multiscale, tseng2005fluent}.
These styles from different levels work together to produce rich expressiveness in speech.
Towards this, some latest researches on 
similar tasks, 
such as emotional speech synthesis \cite{lei2021fine, msemotts} and style transfer \cite{multiscale}, devote effort to performing a multi-scale style modelling which however require auxiliary labels besides text.
To our best knowledge, there is currently no work investigating on multi-scale speaking style prediction just from context.

In this paper, we propose a multi-scale speaking style modelling method to capture and predict multi-scale speaking style from hierarchical context information for expressive TTS.
Our model contains a multi-scale style extractor, a multi-scale style predictor and a FastSpeech 2-based acoustic model.
The extractor is used to extract style embeddings at global level, sentence level and subword level from the ground-truth speech, and to explicitly guide the training of the multi-scale style predictor.
The predictor is based on HCE, 
and we exploit the hierarchical context information of HCE in a more efficient way to predict style embeddings at above three levels.
To reduce the interference or overlapping between speaking styles at different levels, residual style embedding is introduced to represent 
effective style variations in speech. 
Both subjective and objective evaluations on a Mandarin audiobook dataset demonstrate that the proposed method can improve the naturalness and expressiveness of generated speech,
benefiting from its ability to 
accurately 
predict both global-scale and local-scale speaking styles from context.

\section{Methodology}
\begin{figure}[!tb]
	\centering
	\includegraphics[width=0.7\linewidth, height=0.7\linewidth]{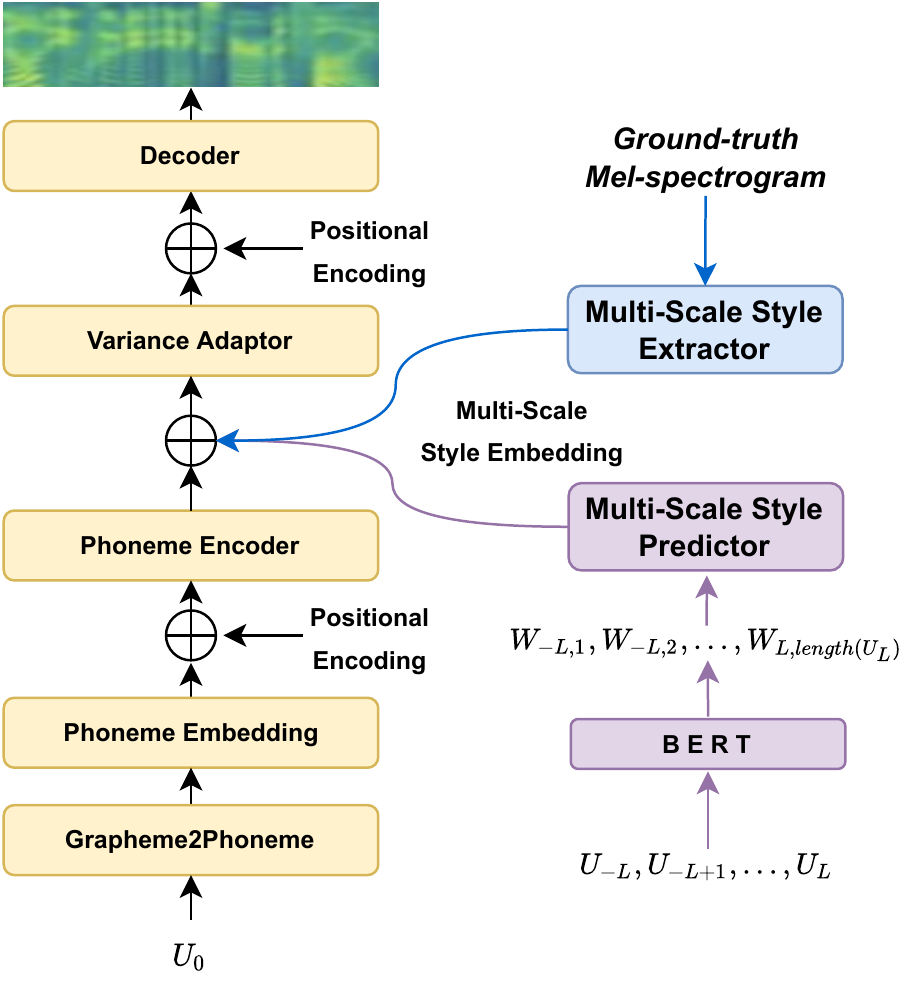}
	\caption{The architecture of our proposed model.}
	\label{fig:architecture}
\end{figure}

The architecture of our proposed model is illustrated in Fig.\ref{fig:architecture}.
It can be mainly divided into three parts, a multi-scale style extractor, a multi-scale style predictor and a FastSpeech 2 \cite{fastspeech2} based acoustic model.
The acoustic model predicts the mel-spectrogram of 
the current sentence
with the assistance of the extractor or the predictor.
The extractor is used to extract the style embeddings at three different levels, and the predictor is used to predict these style embeddings from the context.

\subsection{Multi-Scale Style Extractor}
\begin{figure}[!tb]
	\centering
	\includegraphics[width=0.85\linewidth, height=0.7\linewidth]{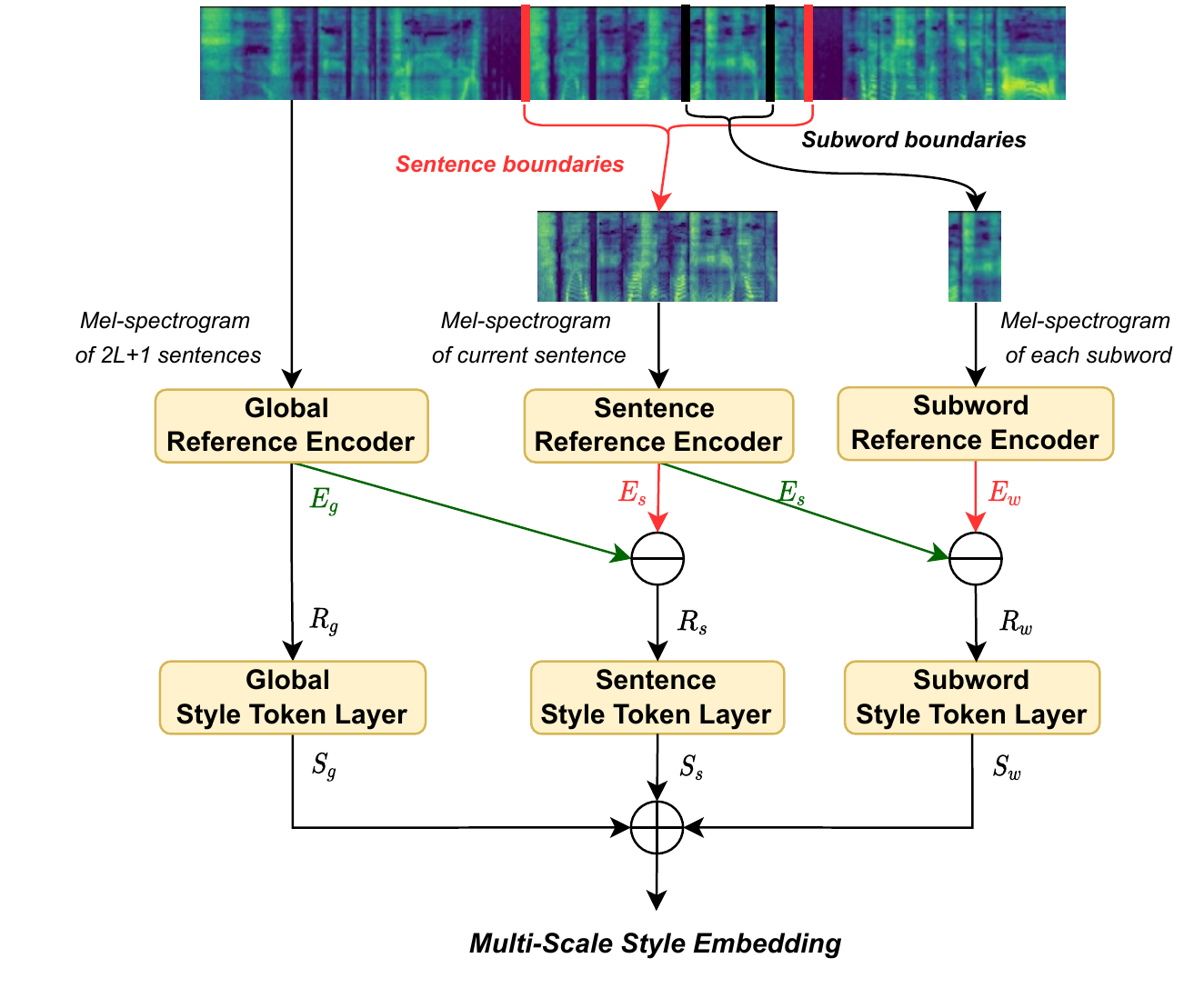}
	\caption{The structure of the multi-scale style extractor.}
	\label{fig:extractor}
\end{figure}

To extract the multi-scale style embedding from the reference speech, we specifically design a multi-scale style extractor, as shown in Fig.\ref{fig:extractor}.
The reference encoders and style token layers corresponding to three different levels make up this module.
All the reference encoders and style token layers have the same architecture and hyperparameters as those of the GST \cite{gst}.

Let $L$ be the number of sentences considered in the past and future context.
The mel-spectrograms corresponding to all $2L+1$ sentences are concatenated and passed to the global reference encoder to extract global reference embedding $E_g$.
The sentence reference encoder is used to extract sentence reference embedding $E_s$ from the mel-spectrogram of the current sentence.
Then the mel-speactrogram of the current sentence is divided by the subword boundaries which are obtained from the forced alignment phoneme boundaries and the subword-to-phoneme alignments.
The mel-spectrogram of each subword goes through the subword reference encoder and the output is denoted as a subword reference embedding $E_w$.
The lower-level embedding $E_w$ may contain redundant style information which has already been covered in the higher-level embedding $E_s$, and similarly for $E_s$ and $E_g$. To reduce such overlapping, 
the residuals between three reference embeddings are represent as the style variation, 
which can be described as:
\begin{align}
    R_g &= E_g \\
    R_s &= E_s-E_g \\
    R_w &= E_w-E_s
\end{align}
where $R_g$, $R_s$ and $R_w$ is the residual style embedding of global-level, sentence-level and subword-level respectively.

The residual reference embeddings are passed to the corresponding style token layers to be decomposed into a fixed number of style tokens respectively, which helps memorize stylistic information at each level and reduce the difficulty of prediction.
After style token layers, the global-level style embedding $S_g$, sentence-level style embedding $S_s$, and subword-level style embedding $S_w$ are obtained.
Finally, for each subword in the current sentence, the
multi-scale style embedding is calculated as the summation of these three levels of style embeddings.

\subsection{Multi-Scale Style Predictor}

To better model the multi-scale speaking style, we extend the HCE in our preliminary work \cite{proposed} and design a multi-scale style predictor.
The lower-level style is derived by being conditioned on the higher-level style. 
This structure is symmetrical with the residual strategy in the style extractor. 

\begin{figure}[!tb]
	\centering
	\includegraphics[width=0.8\linewidth, height=0.4\linewidth]{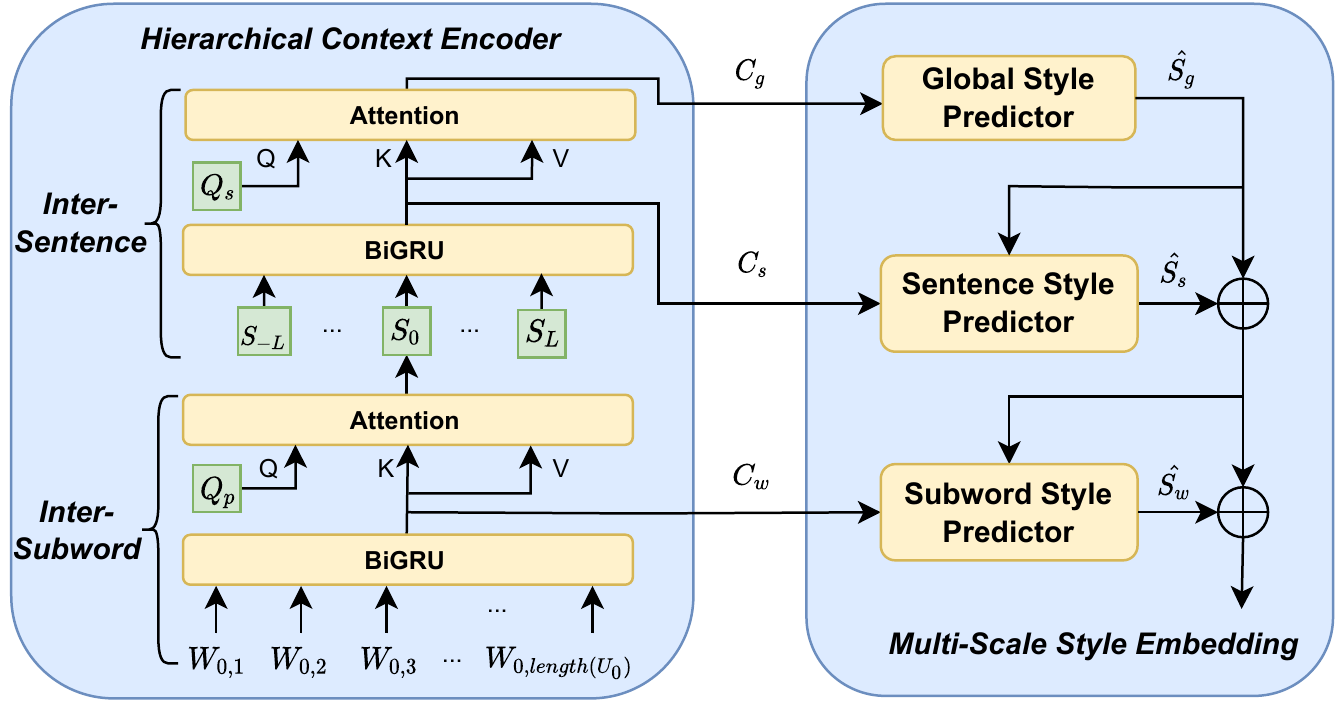}
	\caption{The structure of the multi-scale style predictor.}
	\label{fig:predictor}
\end{figure}

The structure of the multi-scale style predictor is shown in Fig.\ref{fig:predictor}, which consists of the HCE and three extra style predictors.
Each style predictor is composed of a linear layer and Tanh activation.
Besides the current sentence, the multi-scale predictor also considers $L$ sentences in the past or future.
We firstly concatenate all the $2L+1$ sentences to form a long new text, and then pass them to a pre-trained BERT model to obtain the subword-level semantic embedding sequence.
The HCE contains two levels of attention network, the inter-subword, and inter-sentence, each of them contains a bidirectional GRU \cite{gru} and a scaled dot-product attention module \cite{vaswani2017attention}.
The bidirectional GRU is used to get the context embedding by considering temporal relationships, and the attention module is used to aggregate the context embedding sequence into a higher-level embedding.
We denote the output of inter-subword level bidirectional GRU as subword context embedding $C_w$, the output of inter-sentence level bidirectional GRU as sentence context embedding $C_s$, and the output of inter-sentence level attention module as global context embedding $C_g$.

The higher-level style embedding that closer to the global-scale is firstly generated and then utilized as the conditional input to the lower-level style predictor.
In this way, the style embeddings at three levels including $\hat{S_g}$, $\hat{S_s}$, and $\hat{S_w}$ are sequentially generated from the multi-scale style predictor.
The training targets of 
the predictor
come from the corresponding ground-truth style embeddings in the extractor.
It is noteworthy that these three style embeddings of different scales attempt to restore the multi-scale speaking style in the human speech by considering the context information of different levels.
Finally, these embeddings are added together to form the multi-scale style embedding of each subword in the current sentence.

\subsection{Acoustic Model}
As shown in Fig.\ref{fig:architecture}, the backbone of the proposed method is based on FastSpeech 2 \cite{fastspeech2}.
The multi-scale style embedding of each subword in an utterance is provided by the multi-scale style extractor or the multi-scale style predictor.
Then, according to subword-to-phoneme alignments, each multi-scale style embedding is replicated to the phoneme-level and added to the outputs of the phoneme encoder, and then passed to the variance adaptor of FastSpeech 2 for generating mel-spectrogram.

\subsection{Model Training}

Generally, it is challenging for the multi-scale style predictor to derive the multi-scale style in an implicit way.
To encourage this module to learn multi-scale style embedding better, the parameters of our proposed model are trained with the knowledge distillation strategy in three steps.

In the first step, the acoustic model and the multi-scale style extractor are jointly trained to get a well-trained multi-scale style extractor in an unsupervised way.
Especially, to prevent the learning of style on multiple scales from disturbing each other, the reference encoders and style token layer of global-level, sentence-level and subword-level are trained sequentially and separately. When training modules in one of these levels, the rest are frozen. 

In the second step, we leverage knowledge distillation to transfer the knowledge from the multi-scale style extractor to the multi-scale style predictor.
That is, we use the speaking style embedding of multiple levels extracted from the extractor as the target to guide the prediction of the multi-scale style predictor from text.
The loss function of this step is defined as the sum of the mean squared error (MSE) between the extracted speaking style embeddings and the predict speaking style embeddings in global-level, sentence-level, and subword-level.
Finally, we jointly train the acoustic model and multi-scale style predictor with a lower learning rate by considering both the loss of the mel-spectrogram and the loss of the speaking style embedding to further improve the naturalness of synthesized speech.

\section{Experiments}

\subsection{Training Setup}
All the models\footnote{Implemented based on: \href{https://github.com/ming024/FastSpeech2}{https://github.com/ming024/FastSpeech2}\label{fn_fs2}} are trained on an internal Mandarin audiobook dataset, which contains roughly 30 hours of audiobook recordings created by a professional male 
native speaker reading a fiction novel with rich expressiveness.
The dataset has a total of 14,500 audio clips, of which 95\% of the clips are used for training and 5\% of the clips are used for validation and test.

For feature extraction, 80-dimensional mel-spectrograms were extracted with 24kHz sampling rate.
The frame size is set to 1,200 and the hop size is set to 240.
The phoneme duration is extracted by Montreal Forced Aligner \cite{mfa} tool.
In addition, we average ground-truth pitch and energy by duration to get phoneme-level pitch and energy.
An open-source pre-trained Chinese subword-level BERT-base model\footnote{Available at: \href{https://github.com/google-research/bert}{https://github.com/google-research/bert}} is used in our experiments.
The context of current sentence is made up of its two past sentences, itself and its two future sentences.

We train all the models for 220k iterations with a batch size of 16 on an NVIDIA V100 GPU.
For our proposed model, we take 180k iterations to the first train step (60k iterations for each extractor), 20k iterations to the second train step and 20k iterations to the third train step.
The Adam optimizer is adopted with $\beta_1=0.9$, $\beta_2=0.98$, $\epsilon=10^{-9}$ and the warm-up strategy is employed before 4000 iterations.
In addition, we use a well-trained HiFi-GAN \cite{kong2020hifi} as the vocoder to generate waveform.

\subsection{Compared Methods}
To demonstrate the 
performance of our proposed multi-scale model, three 
baseline models are
implemented 
for comparison:

\textbf{FastSpeech 2}
An open-source implementation\textsuperscript{\ref{fn_fs2}} of FastSpeech 2 \cite{fastspeech2}.

\textbf{WSV*}
Word-level style variations (WSV) model.
For a fair comparison,
instead of Tacotron2 \cite{tacotron2} used in the original version of WSV \cite{wsv},
FastSpeech 2 was adopted as the backbone in our implementation.
In addition, an extra bidirectional GRU is used to consider the context information.

\textbf{HCE} 
Hierarchical context encoder (HCE) \cite{proposed} model, which predicts the style on global-level from the context.

\subsection{Subjective Evaluation}
We conduct mean opinion score (MOS) test to evaluate the naturalness and expressiveness of the synthesized speech.
25 native Mandarin speakers are recruited as subjects to rate the given speeches on a scale from 1 to 5 with 1 point interval.
As shown in Table \ref{tab:mos}, the results demonstrate the effectiveness of our proposed methods over the baselines.
There exists a large gap between FastSpeech 2 and Ground Truth, indicating that it is difficult to model multiple speech variations without enough input information.
Our proposed approach achieves the best MOS of 4.058, exceeding FastSpeech 2 by $0.485$, WSV* by $0.377$ and HCE by $0.273$.

\begin{table}[th]\footnotesize
\renewcommand{\arraystretch}{1.0}
  \caption{The MOS on naturalness and expressiveness of different models with 95\% confidence intervals.}
  \label{tab:mos}
  \centering
  \begin{tabular}{l|c} 
    \toprule
    \textbf{Model} &\textbf{MOS} \\
    \midrule
    Ground Truth & $4.665\pm0.074$ ~~~               \\
    FastSpeech 2 & $3.573\pm0.094$ ~~~  \\
    WSV* & $3.681\pm0.080$ ~~~ \\
    HCE & $3.785\pm0.084$ ~~~  \\
    Proposed & $\mathbf{4.058\pm0.074}$ ~~~ \\
    \bottomrule
  \end{tabular}
\end{table}

\begin{figure}[!tb]
	\centering
	\includegraphics[width=0.9\linewidth, height=0.28\linewidth]{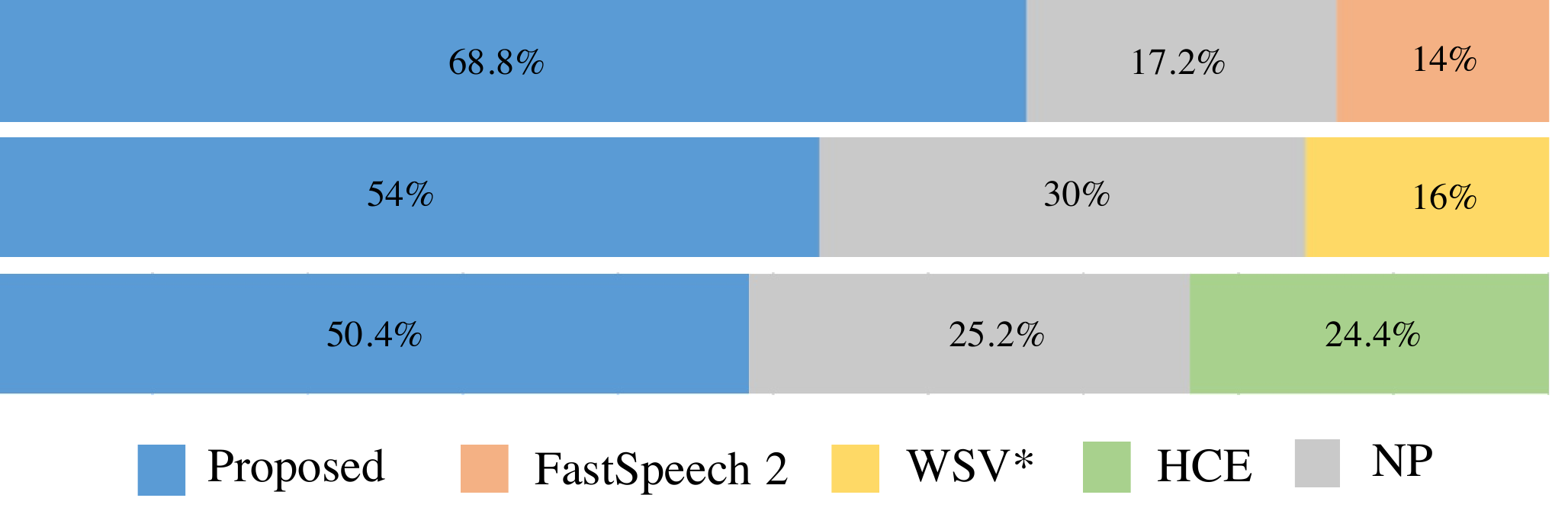}
	\caption{Results of the ABX test on naturalness and expressiveness between different models. NP means no preference.}
	\label{fig:abx}
\end{figure}

ABX preference test is also conducted to ask subjects to give their preferences in terms of naturalness and expressiveness between a pair of speeches generated by different models.
We compare the proposed model with each of the three baseline models.
As shown in Fig.\ref{fig:abx}, the preference rate of our proposed model exceeds FastSpeech 2 by $54.8\%$, WSV* by $38\%$ and HCE by $26\%$ respectively.
Especially, some subjects report the speech synthesized by the proposed model has richer expressiveness than WSV*, and performs better than HCE on the local style properties, such as intonation and stress.

Both MOS and ABX preference tests demonstrate that our proposed approach signiﬁcantly outperforms the three baselines in terms of naturalness and expressiveness.
Compared with the basic FastSpeech 2 that only uses phone sequence as input, the other three models (WSV*, HCE and our proposed model) all perform better, indicating that considering context is helpful for expressive speech synthesis.
Our proposed model also achieves superior performance than not only WSV* that just considers the local-level style, but also HCE that solitarily considers the global-level style.
It demonstrates that modeling the speaking style from different scales can improve the naturalness and expressiveness of synthesized speech significantly.

\subsection{Objective Evaluation}
The root mean square error (RMSE) of F0 and energy, and the MSE of duration are adopted as the metrics of objective evaluation following \cite{wsv, fastspeech2}.
To calculate the RMSE of F0 and energy, we first apply the dynamic time warping (DTW) to construct the alignment paths between the predicted mel-spectrogram and the ground-truth one.
The F0 sequence and energy sequence are then aligned towards ground-truth following the DTW paths.
For the duration, we compute the MSE between the predicted duration and ground-truth duration.
As shown in Table \ref{tab:objective},
our proposed method outperforms the three baselines in all metrics,
which indicates that our proposed model can restore more accurate prosody characteristics, such as pitch, energy and duration, than baselines.
\begin{table}[th]\footnotesize
\renewcommand{\arraystretch}{1}
  \caption{Objective evaluations of different models.}
  \label{tab:objective}
  \centering
  \begin{tabular}{lcccc} 
    \toprule
    \textbf{} & \textbf{F0 RMSE} & \textbf{Energy RMSE} & \textbf{Duration MSE}\\
    \midrule
    FastSpeech 2  & $65.266$ & $5.162$ & $0.2177$~~  \\
    WSV* & $64.807$ & $5.221$ & $0.2051$~~ \\
    HCE  & $63.683$ & $5.045$ & $0.2088$~~  \\
    Proposed  & \textbf{62.544} & \textbf{4.926} & \textbf{0.2014}~~ \\
    \bottomrule
  \end{tabular}
\end{table}

\subsection{Ablation Study}
To demonstrate the effectiveness of several techniques used in our proposed model, including utilizing global-level style, the multi-scale framework and the residual style embedding, we conduct three ablation studies.
Comparison mean opinion score (CMOS) is employed to compare the synthesized speeches in terms of naturalness and expressiveness.
The results are shown in Table \ref{tab:cmos}.
The neglect of modeling global-level speaking style results in $-0.428$ CMOS.
Removing the multi-scale framework (i.e., only modelling the subword-level speaking style) results in $-0.640$ CMOS.
The results indicate the importance of modelling speaking style representation of sentence-level and global-level for expressive speech synthesis.
Moreover, we also find that removing the residual style embedding result in $-0.516$ CMOS.
This indicates that using residual style embedding can represent effective style variations of speech by reducing the interference or overlapping between speaking styles of different levels.

\begin{table}[th]\footnotesize
\renewcommand{\arraystretch}{1.0}
  \caption{CMOS comparison for ablation study.}
  \label{tab:cmos}
  \centering
  \begin{tabular}{l|c} 
    \toprule
    \textbf{Model} &\textbf{CMOS} \\
    \midrule
    Proposed & $0$ ~~~ \\
    \quad -global-level style & $-0.428$ ~~~ \\
    \quad -multi-scale framework & $-0.640$ ~~~ \\
    \quad -residual style embedding & $-0.516$ ~~~ \\
    \bottomrule
  \end{tabular}
\end{table}

\subsection{Case Study}
\begin{figure}[!tb]
	\centering
	\includegraphics[width=1.0\linewidth, height=0.35\linewidth]{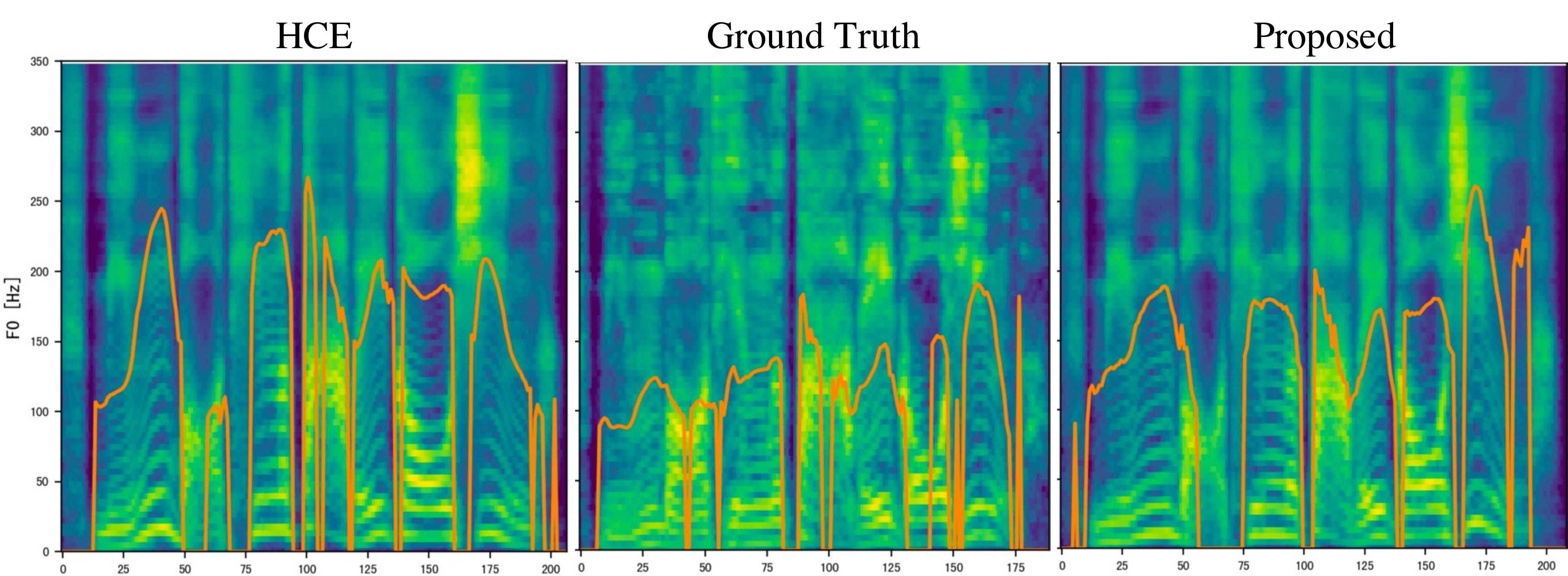}
	\caption{Mel-spectrograms and pitch contours of speeches synthesized by different models for an example utterance in test set.}
	\label{fig:casestudy}
\end{figure}

To explore the impact of the multi-scale speaking style on the expressiveness and naturalness of synthesized speech, a case study is conducted to synthesize an example utterance in test set with HCE and the proposed model, and the ground-truth speech is also provided for reference.
The mel-spectrograms and pitch contours of synthesized speeches as shown in Fig.\ref{fig:casestudy}.
The speech synthesized by HCE contains larger pitch fluctuations than others.
But due to the absence of local-scale speaking style, it lacks the ability to control the local style characteristics of synthesized speech, resulting in a large difference in the trend of intonation compared with ground-truth speech.
Compared with HCE, the speech synthesized by our proposed model is more similar to ground-truth speech in terms of fine-grained style properties, such as the trend of intonation and stress patterns.
With the help of the multi-scale speaking style, our model successfully learns the style variations of human speech, providing the effectiveness of the proposed model.
The result of the case study demonstrates that modelling the multi-scale speaking style from hierarchical context information can effectively improve the naturalness and expressiveness of generated speech.

\section{Conclusions}
In this paper, we propose a multi-scale speaking style modelling method to capture and predict multi-scale speaking style from hierarchical context information for expressive TTS.
Experimental results demonstrate that our proposed method achieves better performance on expressive speech synthesis with the ability to predict both global-scale and local-scale speaking styles from context accurately.

\section{Acknowledgements}

This work is supported by National Key R\&D Program of China (2020AAA0104500), National Natural Science Foundation of China (62076144), National Social Science Foundation of China (13\&ZD189) and Shenzhen Key Laboratory of next generation interactive media innovative technology (ZDSYS20210623092001004).

\newpage

\bibliographystyle{IEEEtran}
\bibliography{references}

\end{document}